\documentclass{article}

\usepackage[nonatbib, preprint]{neurips_2019}

\usepackage{algorithm}
\usepackage[noend]{algpseudocode}
\usepackage{mathtools}

\usepackage{booktabs}
\usepackage{tabularx}
\usepackage{graphicx}
\usepackage{subfig}
\usepackage{pbox}
\usepackage{multirow}
\usepackage[misc]{ifsym}

\usepackage[all]{nowidow}

\usepackage{amsmath}

\DeclareMathOperator*{\argmin}{arg\,min}

\newcommand{\NSR}{NSEEN}
\newcommand{\NSRtitle}{\NSR: Neural Semantic Embedding for \\ Entity Normalization}

\usepackage{titlesec}
\usepackage{wrapfig}

\usepackage[sectionbib, numbers]{natbib}

\usepackage{url}

\usepackage[pagebackref]{hyperref}

\usepackage[dvipsnames]{xcolor}
\definecolor{orange}{RGB}{255,127,0}
\definecolor{cite-green}{RGB}{0,120,0}
\definecolor{cite-blue}{RGB}{0,0,120}
\definecolor{cite-red}{RGB}{120,0,0}

\hypersetup{
    colorlinks, 
    citecolor={cite-green},
    urlcolor={cite-blue},
    linkcolor={cite-red},
    pdfborder={0 0 0},
}

\begin{document}
\title{\NSRtitle}

\author{Shobeir Fakhraei, Joel Mathew, Jos\'{e} Luis Ambite\\
Information Sciences Institute, University of Southern California\\
4676 Admiralty Way, Marina del Rey, CA\\
\texttt{\{shobeir,joel,ambite\}@isi.edu}\\
}

\maketitle              
\begin{abstract}
Much of human knowledge is encoded in text, available in scientific publications, books, and the web. Given the rapid growth of these resources, we need automated methods to extract such knowledge into machine-processable structures, such as knowledge graphs. 
An important task in this process is \textit{entity normalization}, which consists of mapping noisy entity mentions in text to canonical entities in well-known \textit{reference sets}. 
However, entity normalization is a challenging problem; there often are many textual forms for a canonical entity that may not be captured in the reference set, and entities mentioned in text may include many syntactic variations, or errors. 
The problem is particularly acute in scientific domains, such as biology. 
To address this problem, we have developed a general, scalable solution based on a deep Siamese neural network model to embed the \textit{semantic} information about the entities, as well as their \textit{syntactic} variations. 
We use these embeddings for fast mapping of new entities to large reference sets, and empirically show the effectiveness of our framework in challenging bio-entity normalization datasets.  
\end{abstract}


\normalsize
\section{Introduction}
Digital publishing has accelerated the rate of textual content generation to beyond human consumption capabilities.
Taking the scientific literature as an example, Google Scholar has indexed about four and a half million articles and books in 2017 in a 50\% increase from the previous year.  
Automatically organizing this information into a proper knowledge representation is an important way to make this information accessible.  
This process includes identification of entities in the text, often referred to as \textit{Names Entity Recognition (NER)}~\citep{yadav2018survey, mathew2019biomedical}, and mapping of the identified entities to existing reference sets, called \textit{Entity Normalization}, or \textit{Grounding}.
In this paper we propose a text embedding solution for entity normalization to a reference set.

Entity normalization to a reference set is a challenging problem. Even though in some cases normalization can be as simple as a database look-up, often there is no exact match between the recognized entity in the text and the reference entity set. 
There are two main sources for this variation.
The first is \textit{syntactic variations}, where the identified entity contains relatively small character differences with the canonical form present in the reference set, such as different capitalization, reordering of words, typos, or errors introduced in the NER process (e.g., `FOXP2' and `FOX-P2').  
The second and more challenging problem,  which we call \textit{semantic variations},  is when the identified entity does not exist in the reference set, even when considering significant syntactic variations, but a human reader can recognize the non-standard entity name. For example, entities often have multiple canonical names in the reference set and the identified entity name is a combination of parts of different canonical names (e.g., `P70~S6KA' and `52 kDa ribosomal protein S6 kinase').

A further challenge is how to perform normalization at scale. 
Exhaustive pairwise comparison of the identified entity to the reference entities grows quadratically and is unfeasible for large datasets. 
\textit{Blocking}~\citep{papadakis2016comparative} techniques speed up the process by selecting small subsets of entities for pairwise comparisons. 
Unfortunately, blocking methods applied directly to the textual representation of the entity names are often limited to simple techniques that can only address syntactic variations of the entity names. So, traditional blocking may eliminate matches that are semantically relevant but syntactically different. 

To address these issues, we develop a text embedding solution for entity normalization. 
Our contributions include:  
(1)~A general, scalable deep neural-based model to embed entity information in a numeric vector space that captures both \textit{syntactic} and \textit{semantic variations}.
(2)~An approach to incorporate syntactic variations of entity names into the embeddings based on domain knowledge by extending the use of contrastive loss function with soft labels.
(3)~A method for dynamic hard negative mining to refine the embedding for improved performance.
(4)~Using an approximate \textit{k-nearest neighbors} algorithm over the embeddings to provide a scalable solution without the need for traditional blocking.
\section{Related Work}

\textit{Data Normalization}, linking entities to their canonical forms, is one of the most fundamental tasks in information retrieval and automatic knowledge extraction~\citep{christen2012data}. 
Many related tasks share components with entity normalization, but also have subtle differences. 
Record linkage~\citep{koudas2006record}, aims to find records from different tables corresponding to the same entity. Records often contain multiple fields and one of the challenges in this task is reasoning on different fields, and their combinations.
Deduplication~\citep{elmagarmid2007duplicate} is similar to record linkage, but focuses on the records of the same table, so it does not have to consider the heterogeneity of fields across different tables.
Entity resolution~\citep{getoor2012entity}, is a more general term that deals with findings entity mentions that refer to the same entity and often inferring a canonical form from them. 

A critical feature in our setting is the presence of a canonical \textit{reference set}, so that we ask  
``which canonical entity a mention is mapped to?'' in contrast to ``which records are the same?'' for settings were the canonical entity is latent. Reference sets are specially important in bio-medical domains~\citep{leaman2016taggerone}. Unlike record linkage, we do not have multiple fields and only reason on a single string. 

Feature-engineered string similarities~\citep{cheatham2013string} form the core of most traditional entity resolution methods. In contrast, Our approach learns a \textit{similarity metric} for entity normalization based on syntactic and semantic information. We compute these similarities via embedding the entity mentions into a vector space.
Text embeddings, such as word2vec~\citep{mikolov2013distributed}, GloVe~\citep{pennington2014glove}, or more recently ELMo~\citep{Peters:2018}, and BERT~\citep{devlin2018bert} have been very successful in language processing and understanding applications, in great measure because they have been computed over very large corpora.
However, these methods are not task specific and provide general embeddings based on the text context. 
Our approach is based on computing direct similarities rather than analyzing the surrounding text. 
Hence, for Entity Normalization, we use a deep Siamese neural network that has been shown to be effective in learning similarities in text~\citep{neculoiu2016learning} and images~\citep{taigman2014deepface}. Both of these approaches define a contrastive loss functions~\citep{hadsell2006dimensionality} to learn similarities.  
Recently, \citet{joty2018distributed} and \citet{mudgal2018deep} proposed deep neural network methods for record linkage (with multiple fields) in a database. A major focus of their work was on combining data in different fields. Our setting differs since we operate on entity name strings, and match them to canonical references. 

To avoid exhaustive pairwise computation of similarities between entities often blocking~\citep{michelson2006learning} or indexing~~\citep{christen2012survey} techniques are used to reduce the search space. These methods are often based on approximate string matching. The most effective methods in this area is based on hashing the string with the main purpose of blocking the entities as a pre-processing step, followed by the matching part that is performed after blocking. 
In our method, we combine both steps by mapping the entity mentions to a numerical space to capture similarities. The blocking in our case conceptually follows the matching process via applying approximate nearest neighbors approaches on our semantic embedding space.

In the biomedical domain, \citet{kang2012using} propose a rule-based method and \citet{leaman2013dnorm} propose a learning-to-rank-based approach for disease normalization. \citet{leaman2016taggerone} further perform joint name entity recognition and normalization. We provide an embedding-based approach for entity normalization. We perform our experimental validation on two biomedical datasets of protein and chemical entities. 
\section{Approach}

\textbf{Problem Definition:~}
Given a query entity mention ($n_q$), and a reference set of entities $\mathcal{R} \equiv \{e_1, \dots, e_m \}$, where each entity $e_i \equiv < \lambda_i, \{ n_i^1, \dots, n_i^k \} >$ is identified via an ID ($\lambda_i$) and an associated set of names ($n_i^k$) that refer to the entity, our goal is to return the ID ($\lambda_q$) of the corresponding entity in our reference set $\mathcal{R}$. 
The exact textual name of the query entity may not exist in the reference~set.

We map this normalization task to an approximate nearest-neighbors search in a n-dimensional space where each name $n_l^m$ in the reference set is encoded into a numerical vector representation $v_l^m$. 
Our objective in this embedding space is that names of the same entity (even syntactically very different) be closer to each other compared to names of other entities. That is, ${n_l^m \to v_l^m}$ such that ${\delta(v_l^m, v_l^p) < \delta(v_l^m, v_o^*)}$, where $e_l$ and $e_o$ are entities ${(e_l \neq e_o)}$, $n_*^*$ their corresponding names, $v_*^*$ embedding vectors of these names, and $\delta$ is a distance function.

We use a Siamese neural network architecture to embed the semantic information about the entities as well as their syntactic similarities. 
We further refine the similarities via dynamic hard negative sampling and incorporating domain knowledge about the entities using additional generated training data. We then encode and store the embeddings in a numeric representation that enables fast retrieval of the results without the need for traditional character-based blocking.
Our approach consist of three steps:

\noindent \textbf{Similarity Learning.} We first learn an embedding function ($\mathcal{M} : n \to v$) that maps the entity names to a numeric vector space where names of the same entities are close to each other. 

\noindent \textbf{Embedding and Hashing.} Then, we embed all the names in the reference set $\mathcal{R}$ to the numerical vector space and hash and store the reference set embeddings for fast retrieval.

\noindent \textbf{Retrieval.} Finally, we embed the query name (i.e., $n_q \to v_q$) using the learned model $\mathcal{M}$ and find the closest samples to it in the embedding space to retrieve the corresponding ID ($\lambda_q$) of the query name in the reference set.  

\noindent The following sections describe each step in detail.

\begin{figure}[t]%
    \centering
    \subfloat[Overall Framework\label{fig:training_framework}]{\includegraphics[width=0.46\columnwidth]{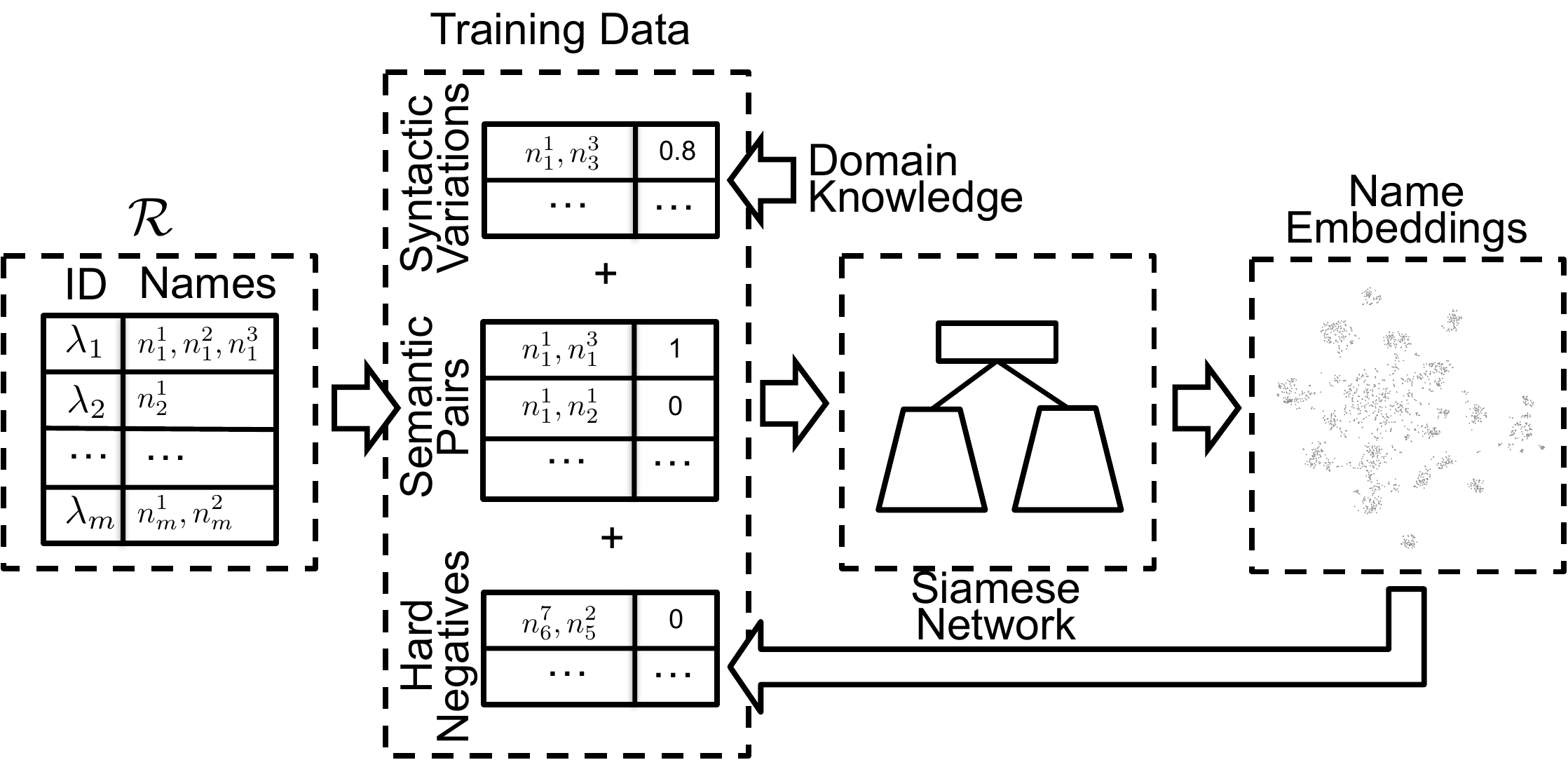}}%
    \qquad
    \subfloat[Siamese Network\label{fig:siamese_network}]{\includegraphics[width=0.46\columnwidth]{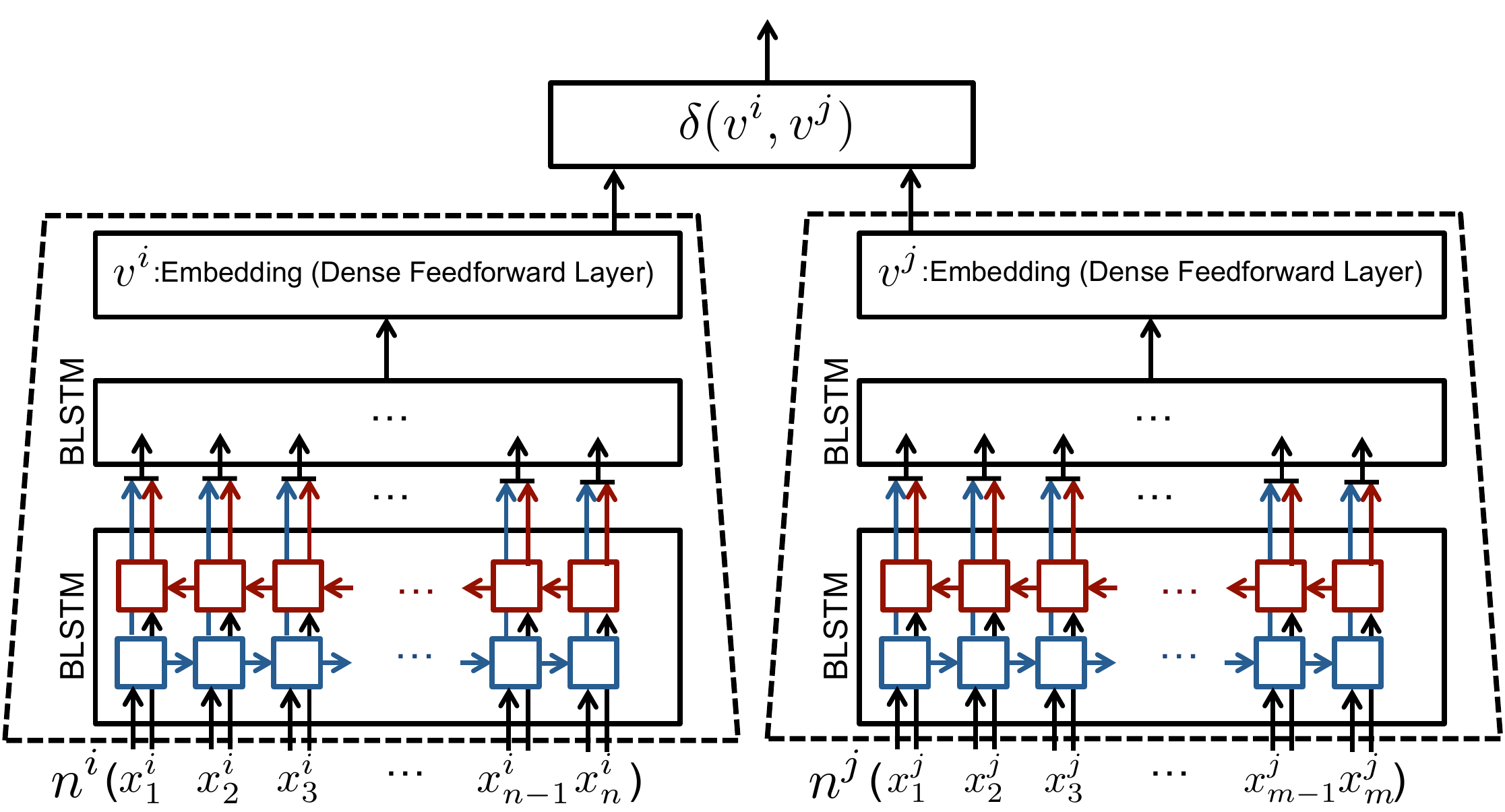} }%
    \caption{Learning embedding function based on semantics in reference set and syntactic variations defined by domain knowledge and hard negative mining.}%
\end{figure}

\subsection{Similarity Learning}

We first learn a function ($\mathcal{M}$) that maps the textual representation of entity names ($n$) to a numerical vector representation ($v$) that preserves the proximity of names that belong to the same entity, using a Siamese recurrent neural network model.  Figure~\ref{fig:training_framework} shows the overall approach and Algorithm~\ref{alg:training} describes the similarity learning process. 

\begin{algorithm}
\caption{\NSR: Similarity Learning}
\label{alg:training}
\begin{algorithmic}[1]
\Procedure{TrainSim}{$\mathcal{R},P_d$}
\State{\textbf{Input:} $\mathcal{R}$ reference set}
\State{\textbf{Input:} $P_d$ pairs based on knowledge of syntactic variation in the domain}
\State Generate pairs based on reference set $\mathcal{R}$ and add them to training data $\mathcal{D}$
\State Add $P_d$ pairs to the training data $\mathcal{D}$
\For{k times}
\State Train the model $\mathcal{M}$ (Siamese network) on $\mathcal{D}$
\State Embed all the names in $\mathcal{R}$: $n \to v$
\ForAll{$v_l^i$}  \Comment{Hard negative mining}
\State find the k closest $v_k^j$ to $v_l^i$
\If{$k \neq l$} 
\State add $<n_k^j,n_l^i,0>$ to training data $\mathcal{D}$
\EndIf
\EndFor
\EndFor
\State \textbf{return} $\mathcal{M}$\Comment{The trained embedding function}
\EndProcedure
\end{algorithmic}
\end{algorithm}

\subsubsection{Siamese Recurrent Neural Network}~\\
\label{sec:network}
The Siamese neural network architecture of two towers with shared weights and a distance function at the last layer has been effective in learning similarities in domains such as text~\citep{neculoiu2016learning} and images~\citep{taigman2014deepface}. 
Figure~\ref{fig:siamese_network} depicts an overview of the network used in our framework. 

We feed pairs of names and a score indicating the similarity of the pairs (i.e., $<n^i,n^j,y>$) to the Siamese network. As shown in Figure~\ref{fig:siamese_network}, $n^i$ and $n^j$ are entity names represented as a sequences of characters $<x^i_1,\dots,x^i_n>$ and $<x^j_1,\dots,x^j_m>$, and $y \in [0, 1]$ represents the similarity between the names. 
To read the character sequence of the names, we feed the character embedding to four layers of Bidirectional-LSTM, followed by a single densely connected feed-forward layer, which generate the embeddings $v$.

\noindent {\bf Contrastive Loss Function.}
While we can use several distance functions ($\delta$) to compare the learned vectors of the names, we use cosine distance between the embeddings $v_i$ and $v_j$, due to its better performance in higher dimensional spaces.
We then define a contrastive loss \citep{hadsell2006dimensionality} based on the distance function $\delta$ to train the model, as shown in equation~\ref{eq:loss_func}. The intuition behind this loss function is to pull the similar pairs closer to each other, and push the dissimilar pairs up to a margin $m$ apart ($m=1$ in our experiments).

\begin{equation}\label{eq:loss_func}
    \mathcal{L} = \frac{1}{2} y \delta(v_i,v_j)^2 + \frac{1}{2} (1-y) \max(0,m-\delta(v_i,v_j))^2
\end{equation}

The contrastive loss has been originally proposed for binary labels where we either fully pull two points towards each other or push them apart. 
In this paper, we propose to extend this loss via using soft real-valued labels when we introduce syntactic variations of the names described in section~\ref{sec:pairgeneration} to indicate uncertainties about the similarities of two vectors.
For the margin of 1 (i.e., $m=1$), the distance that minimizes the loss function $\mathcal{L}$ for the real-valued label $y$ is:%
\footnote{for brevity of notation we denote $\delta(v_i,v_j)$ with $\delta_v$}

\begin{align}\label{eq:loss_real_value}
\begin{split}
    \frac{\partial \mathcal{L}}{\partial \delta_v} &= y \delta_v - (1-y) (1-\delta_v)\\
    \argmin_{\delta_v}\mathcal{L} &= \{\delta_v \mid y + \delta_v - 1 = 0 \} = 1 - y 
\end{split}
\end{align}

\begin{figure}
{
\centering
\includegraphics[width=0.5\columnwidth]{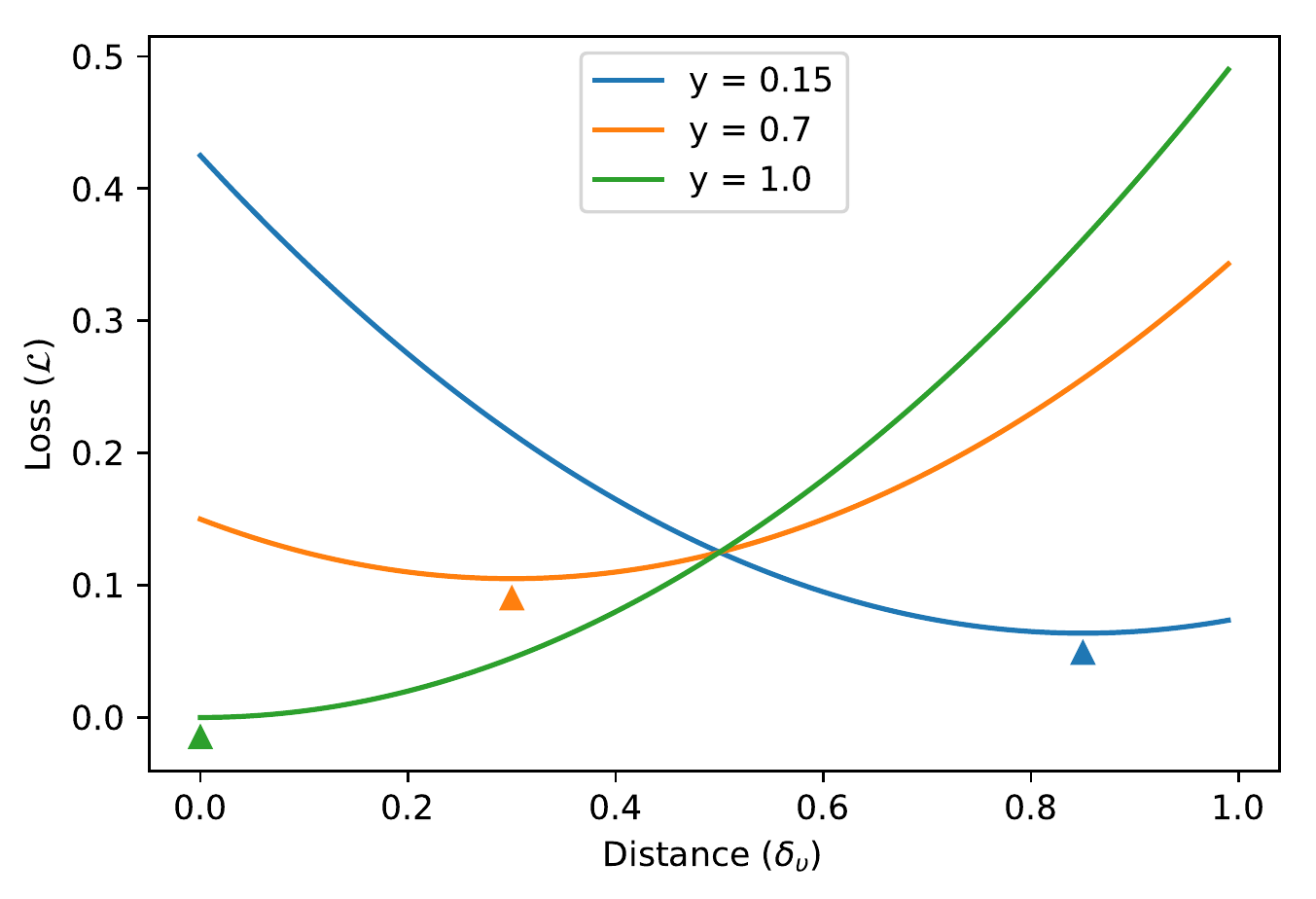}
\caption{Contrastive loss ($\mathcal{L}$) based on distance values ($\delta_v$) for different real-value labels $y$.(Best viewed in color)}
\label{fig:contrastive_loss}
}
\end{figure}

For example, in our setting the optimal distance between the embeddings of two names with 0.7 similarity (i.e., $y=0.7$) is 0.3. 
Figure~\ref{fig:contrastive_loss} depicts the changes in loss when altering the distance corresponding to different $y$ values, and the points that minimize the loss (i.e., $\argmin_{\delta_v}\mathcal{L}$) are marked on each line.  

\subsubsection{Pair Selection and Generation}~\\
\label{sec:pairgeneration}
In order to train the model we need labeled pairs of names ($<n^i,n^j,y>$). 
We generate three sets of pairs using different approaches: (1)   
the \textit{initial set} based on the names in the reference set, (2) the \textit{syntactic variations set} based on domain knowledge, and (3) the \textit{hard negative set}.
The initial and the hard negative pairs capture the \textit{semantic} relationships between names in the reference set, and the syntactic variations capture the \textit{syntactic} noise that may be present in referring to these names in reality.

\noindent \textbf{Initial Semantic Set.}
We generate an initial training set of similar and dissimilar pairs based the entities in the reference set $\mathcal{R}$.
We generate positive pairs by the cross product of all the names that belong to the same entity, and initialize the negative set of dissimilar pairs by randomly sampling names that belong to different entities. Formally: 
\begin{align*}
    \begin{split}
    P_+ &= \{{<n^i,n^j,1>} ~\mid~ (\forall n^i_l, n^j_l \in e_l) \wedge (\forall e_l \in \mathcal{R})\}\\
    P_- &= \{{<n^i,n^j,0>} ~\mid~ (n^i_l, n^j_m \in e_l,e_m) \wedge (e_l,e_m \in \mathcal{R}) \wedge (e_l\neq e_m)\}
    \end{split}
\end{align*}

\noindent \textbf{Syntactic Variations and Entity Families.}
In order to train the model with the syntactic variations that could be introduced in the real-world textual representation of the names, we add pairs of names to the training set and label them with their real-value string similarities. 
The argument behind using real-valued labels is provided in equation~\ref{eq:loss_real_value}, with the intuition that using a label of 0 will completely repel two vectors and using a label of 1 will bring two vectors as close as possible, but using a label between 0 and 1 will aim to keep the two vectors somewhere inside the margin.

We use Trigram-Jaccard, Levenshtein Edit Distance, and Jaro–Winkler to compute string similarity scores \citep{cohen2003comparison} between the pairs of names and include sets of pairs with labels based on each similarity score in the training set. 
The intuition is that the model will learn a combination of all these string similarity measures. 
To select the name pairs to include in this process, we consider two sets of variations based on the \textit{same name}, and \textit{different names}. 

\textit{Same name} variations are the noise that can be introduced to an extracted name in real-world settings. To capture the most common forms of noise occurring on the same name, we make the following three modifications based on our observation of the most frequent variations in the query names:
\begin{itemize}
\item Removing the spaces, e.g., $<$FOX P2, FOXP2, \textit{y}$>$
\item Removing all but alphanumerical characters, e.g., $<$FOX-P2, FOXP2, \textit{y}$>$
\item Converting to upper and lower cases, e.g., $<$Ras, RAS, \textit{y}$>$, $<$Ras, ras, \textit{y}$>$
\end{itemize}

\textit{Different name} variations introduce a higher level of similarity concept to the model. We make the second set of pairs by selecting the names of entities that are \textit{somehow related} and computing their string similarities. 
For example, in our experiments with proteins we select two entities that belong to the same protein family and generate pairs of names consisting of one name from each. 
The labels are assigned to these pairs based on their string similarities.
This set of pairs not only introduces more diverse variations of textual string similarities, it also captures a \textit{higher-level relationship} by bringing the embeddings of the names that belong to a group closer to each other. 
Encoding such hierarchical relations in the entity representations has been effective in various domains~\citep{chen2018harp}. 

\noindent \textbf{Hard Negative Mining.}
Given the large space of possible negative name pairs (i.e., the cross product of the names of different entities) we can only sample a subset to train our model. 
As stated earlier we start with an initial random negative sample set for our training. 
However, these random samples may often be trivial choices for the model and after a few epochs may not contain enough useful signal.
The use of contrastive loss makes this issue more problematic as the probability of the distance between randomly selected negative samples being less than the margin ($m$) is low.  
Sampling techniques, often called \textit{hard-negative mining}, have been introduces in domains such as knowledge graph construction~\citep{kotnis2017analysis} and computer vision~\citep{shrivastava2016training} to deal with similar issues. 

The idea behind hard negative mining is finding negative examples that are most informative for the model. These examples are the ones closest to the decision boundary and the model will most likely assign a wrong label to them. 
As shown in Figure~\ref{fig:training_framework} and Algorithm~\ref{alg:training}, we find the hard negatives by first embedding all the names in the reference set~$\mathcal{R}$ using the latest learned model~$\mathcal{M}$. We then find the closest names to each name in the embedding space using an approximate k-nearest neighbors algorithm for fast iterations. 
We then add the name pairs found using this process that do not belong to the same entity with a 0 label to our training set and retrain the model $\mathcal{M}$. We repeat this process multiple times to refine the model with several sets of hard negative samples. 

\subsection{Reference Set Embedding and Storage}

The model $\mathcal{M}$ that we trained in the previous step is basically a function that maps a name string to a numerical vector. Since both towers of the Siamese network share all their weights, the final embedding is independent of the tower the original string is provided to as input. 
Considering the goal of our framework, which is to perform entity normalization of query names
($n_q$) 
to the entities in the reference set $\mathcal{R}$, we embed all the names in the reference set using the final trained model $\mathcal{M}$, and store the embeddings for comparison with future queries. 

Our task becomes assigning an entity in our reference set to the query name $n_q$ by finding the closest entity to it in the embedding space. 
This assignment is basically a nearest neighbor search in the embedding space. The most naive solution to this search would entail a practically infeasible task of exhaustive pairwise comparisons of query embedding with all embeddings in a potentially large reference set.  
Moreover, since we iteratively repeat the nearest neighbor look-up in our training process for hard-negative mining, we need a faster way to retrieve the results.

This challenge is prevalent in many research and industry applications of machine learning such as recommender systems, computer vision, and in general any similarity-based search, and has resulted in development of several fast \textit{approximate nearest neighbors} approaches~\citep{ponomarenko2014comparative, rastegari2013predictable}. 
We speed-up our nearest neighbors retrieval process by transforming and storing our reference set embeddings in an approximate nearest neighbors data structure. Algorithm~\ref{alg:embedding} describes the overall process of this stage. 

We leverage a highly optimized solution that is extensively used in applied settings, such as Spotify, to deal with large scale approximate nearest neighbor search, called \textit{Annoy (Approximate Nearest Neighbors Oh Yeah!)}~\citep{bernhardssonannoy}.
Annoy, uses a combination of random projections and a tree structure where intermediate nodes in the tree contain random hyper-planes dividing the search space. It supports several distance functions including Hamming and cosine distances based on the work of \citet{bachrach2014speeding}.

Since we have already transformed the textual representation of an entity name to a numerical vector space, and the entity look-up to a nearest neighbor search problem, we can always use competing approximate nearest neighbors search methods~\citep{naidan2015non}, and the new state-of-the-art approaches that will be discovered in the future. 
Furthermore, using such scalable data structures for our embeddings at this stage preserves semantic similarities learned by our model, in contrast to traditional blocking approaches applied as a pre-processing step that could break the semantic relationship in favor of textual similarities. 

\subsection{Retrieval}
During the retrieval step, depicted in Algorithm~\ref{alg:retrieval} we first compute an embedding for the query name based on the same model $\mathcal{M}$ that we used to embed the reference set. 
We then perform an approximate nearest neighbor search in the embedding space for the query name, and return the ID of retrieved neighbor as the most probable entity ID for the query name. 
Note that in our setup we do not need to perform a separate direct look up for the query names that \textit{exactly} match one of canonical names in the reference set. If the query name is one of the canonical names in the reference set, it will have exactly the same embedding and zero distance with one of the reference set names. 

\begin{center}
\noindent
\begin{minipage}[htbp]{6cm}
\begin{algorithm}[H]
\caption{Embedding $\mathcal{R}$}
\label{alg:embedding}
\begin{algorithmic}[1]
\Procedure{Embed}{$\mathcal{R},\mathcal{M}$}
\ForAll{$n_i \in \mathcal{R}$} 
\State $n_i \xrightarrow[]{\mathcal{M}} v_i$
\EndFor
\ForAll{$v_i$}
\State Hash $v_i$ and store in $\mathcal{H}_{v_i}$
\EndFor
\State \textbf{return} $\mathcal{H}_v$\Comment{Hashed embeddings}
\EndProcedure
\end{algorithmic}
\end{algorithm}
\end{minipage} ~~
\begin{minipage}[htbp]{6cm}
\begin{algorithm}[H]
\caption{Retrieval}\label{alg:retrieval}
\begin{algorithmic}[1]
\Procedure{Retrieve}{$\mathcal{H}_v$, $\mathcal{M}$, $n_q$}
\State Embed the query name: $n_q \xrightarrow[]{\mathcal{M}} v_q$
\State Find the closest $v_k^j$ to $v_q$ using approximate nearest neighbor search (Annoy) on  $\mathcal{H}_v$ 
\State \textbf{return} $\lambda_k$ as the ID (i.e., $\lambda_q$) 
\EndProcedure
\end{algorithmic}
\end{algorithm}
\end{minipage}
\end{center}
\section{Experimental Validation}
We conduct two set of experiments mapping query names to their canonical names to empirically validate the effectiveness of our framework. 
The two references sets are \textit{UniProt} for proteins and \textit{ChEBI} for chemical entities, and the query set is from PubMed extracts provided by the BioCreative initiative~\citep{arighi2017bio}, as detailed in the following sections.

\subsection{Reference Sets}
The reference sets we use in our experiments are publicly available on the internet, and are the authority of canonical entity representations in their domains.

\begin{description}
\item[UniProt.] The Universal Protein Resource (UniProt) is a large database of protein sequences and associated annotations~\citep{apweiler2004uniprot}. For our experiments, we use the different names associated with each human protein in the UniProt dataset and their corresponding IDs. Hence, the task here is mapping a human protein name to a canonical UniProt ID.
\item[ChEBI.] We used the chemical entity names indexed in the Chemical Entities of Biological Interest (ChEBI) ontology. 
ChEBI is a dataset of molecular entities focused on ‘small’ chemical compounds, including any constitutionally or isotopically distinct atom, molecule, ion, ion pair, radical, radical ion, complex, conformer, identifiable as a separately distinguishable entity~\citep{hastings2015chebi}. The task here is mapping a small molecule name to a canoncal ChEBI ID.
\end{description}

\begin{figure}[b]
    \centering
    \subfloat[UniProt]{{\includegraphics[width=0.45\columnwidth,trim={0.4cm 0.1cm 1.5cm 1.2cm},clip]{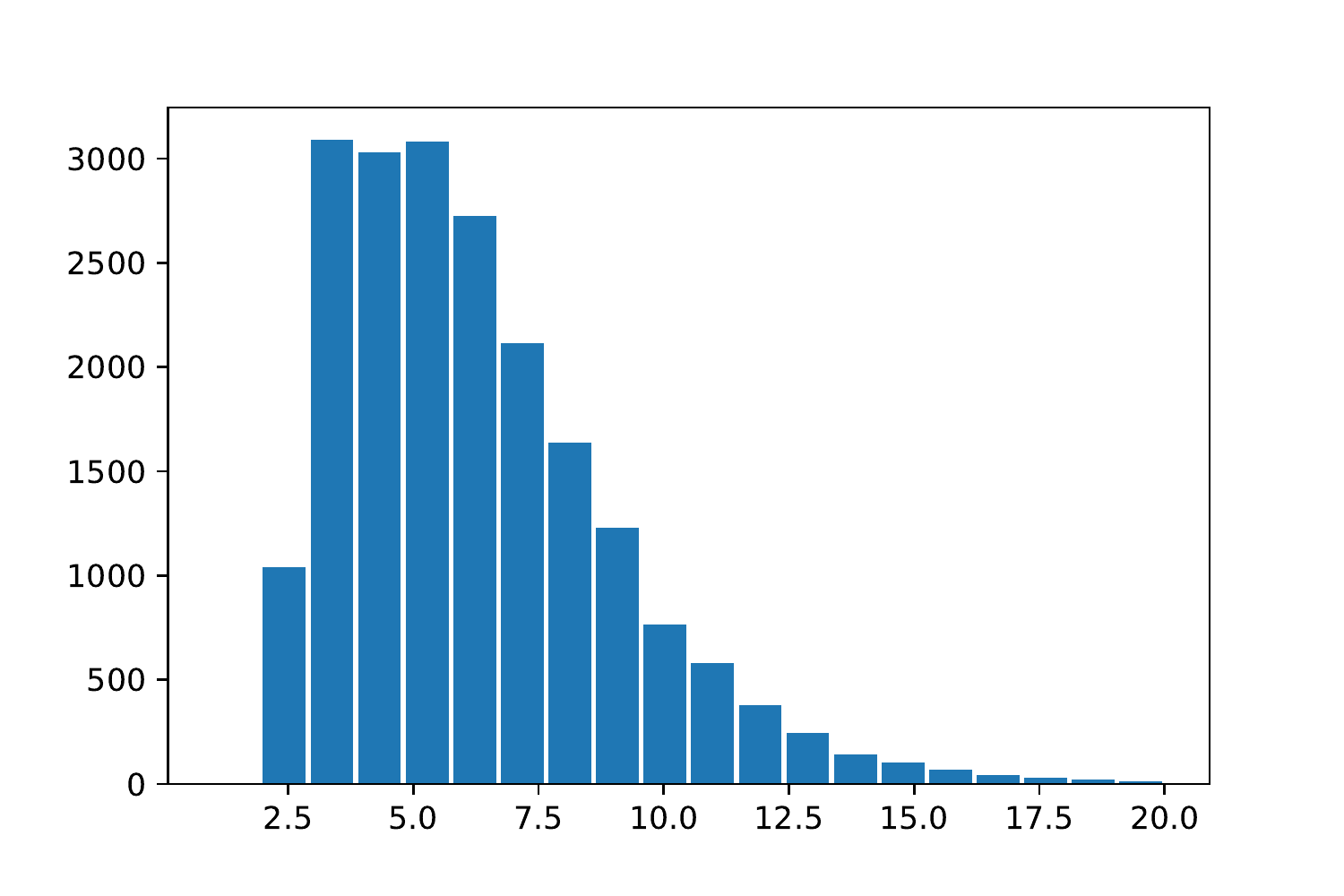}}}%
    \qquad
    \subfloat[ChEBI]{{\includegraphics[width=0.45\columnwidth,trim={0.4cm 0.1cm 1.5cm 1.2cm},clip]{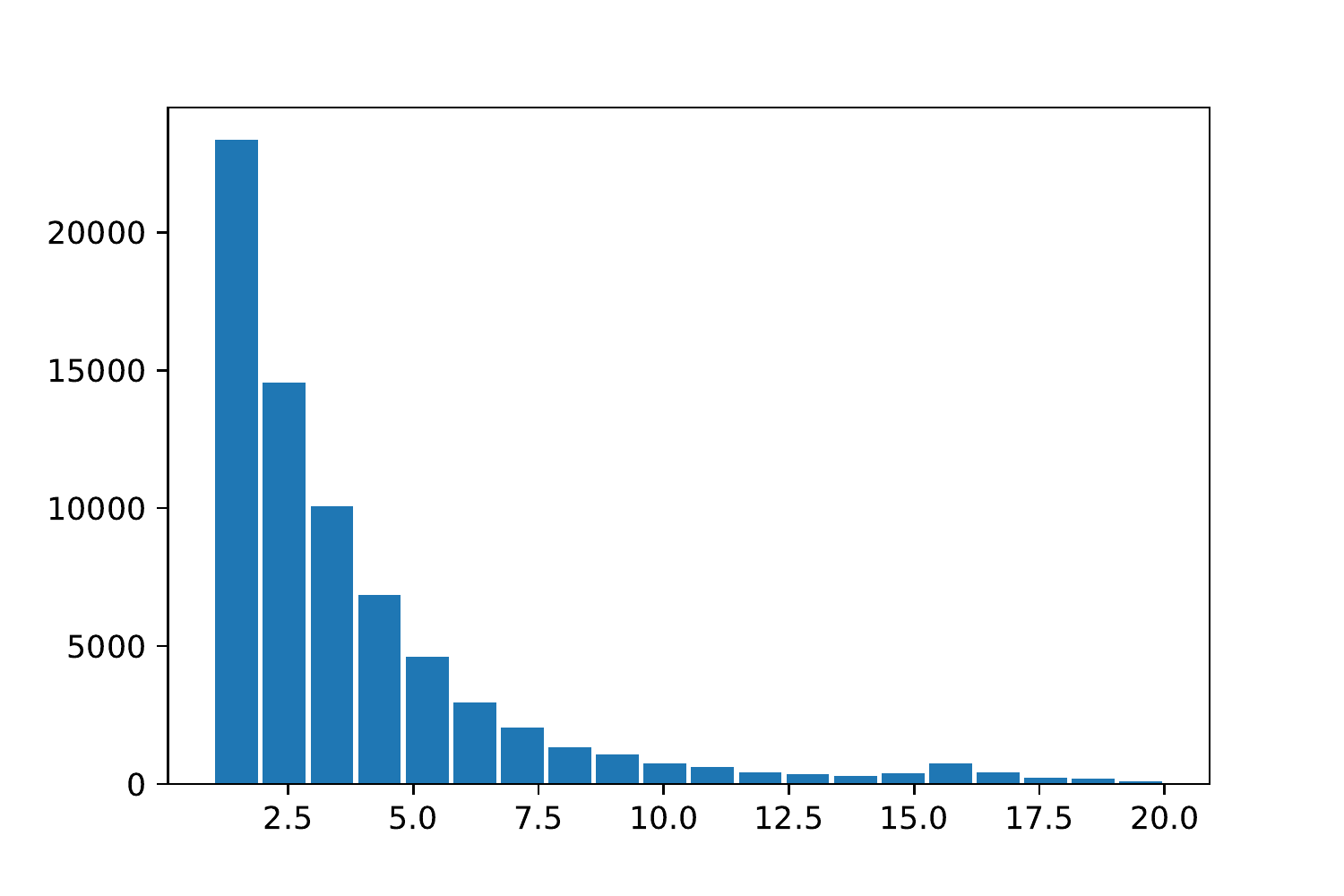}}}%
    \caption{Distribution of names per entity in the reference datasets.}
    \label{fig:refset_hist}%
\end{figure}

Table~\ref{tab:referenceset_stats} depicts the total number of entities ($e_i$) and their corresponding ID--name pairs (${<\lambda_i, n_i^j>}$) in the reference sets, showing UniProt having less number of entities, but more names per entity comparing to ChEBI. 
Moreover, Figure~\ref{fig:refset_hist} depicts the histogram that shows the distribution of the number of names per each entity in the reference sets. 
Note that there are no entities in the UniProt reference set with only one name, but there are many proteins with several names.
In contrast, the ChEBI dataset contains many entities with only one name.

\begin{table}[htbp]
\centering
\caption{Statistics of the entities in the reference sets}
\label{tab:referenceset_stats}
    \begin{tabularx}{\hsize}{X X X}
    \toprule
    \textbf{Datasets} & \textbf{entities} & \textbf{$<$entity, name$>$ pairs} \\
    \midrule
    UniProt (Human)  & 20,375 & 123,590\\
    ChEBI  & 72,241 & 277,210 \\
    \bottomrule
    \end{tabularx}
\end{table}

\subsection{Query Set}
We use the datasets provided by the BioCreative VI Interactive Bio-ID Assignment Track~\citep{arighi2017bio} as our query data. These datasets provide several types of bio-medical entity annotations generated by SourceData curators that map published article texts to their corresponding database IDs. 
The main interesting point about the BioCreative corpus for entity normalization is that the extracted entity names come from published scientific articles, and contain the entity-name variations and deviations forms that are present in the real world. 

The Bio-ID datasets include a separate \textit{train} and \textit{test} sets. 
We use both of these datasets as query sets with gold standard labels to evaluate our method. 
The training set (we name it BioC1) consists of 13,573 annotated figure panel captions corresponding to 3,658 figures from 570 full length articles from 22 journals, for a total of 102,717 annotations. 
The test data set (we name it BioC2) consisted of 4,310 annotated figure panel captions from 1,154 figures taken from 196 full length journal articles, with 30,286 annotations in total \citep{arighi2017bio}. 

Table~\ref{tab:bioc_stats} shows the number of UniProt and ChEBI entities in the annotated corpus.
In our experiments we keep the original training (BioC1) and test (BioC2) splits of the data for reproducablility and ease of future comparisons, but we should note that for our purposes both BioC1 and BioC2 are just a source of correct normalizations with gold standards, and test sets in our experiments. Our algorithm is not trained on any of these datasets. 

\begin{table}[htbp]
\centering
\caption{Statistics of the annotations in the BioCreative~VI Bio-ID corpus}\label{tab:bioc_stats}
    \begin{tabularx}{\hsize}{@{}c X X X X X X@{}}
    \toprule
    \textbf{Dataset} & & \multicolumn{2}{c}{\textbf{UniProt}} & & \multicolumn{2}{c}{\textbf{ChEBI}}\\
    \cline{3-4}
    \cline{6-7}
    ~ & & \textbf{Mentions} & \textbf{Entities} & & \textbf{Mentions} & \textbf{Entities}\\
    \midrule
    BioC1 & & 30,211 & 2,833 & & 9,869 & 786\\
    BioC2 & & 1,592 & 1,321 & & 829 & 543\\
    \bottomrule
    \end{tabularx}%
\end{table}

\subsection{Baselines}

\noindent {\bf USC--ISI.}
As a representative of traditional record linkage techniques, we use the current production system for Named Entity Grounding at USC Information Science Institute, developed for the DARPA Big Mechanism program, as one of the baselines. 
The system is an optimized solution that employs a tuned combination of several string similarities including Jaccard, Levenshtein, and JaroWinkler distances with a prefix-based blocking system. 
It also includes a post re-ranking of the results based on the domain knowledge, such as the curation level of the entity (e.g., if the protein entry in UniProt has been reviewed by a human or not), the matching between the ID components and the query name, and popularity of the entities in each domain. 
This system provides entity grounding for several bio-medical entities including Proteins and Chemicals, and is publicly available at~\citep{ambite2017}. 
The system can produce results based on the FRIL~\citep{jurczyk2008fril} record linkage program and Apache Lucene~\citep{bialecki2012apache},
and we use the overall best results of both settings as the baseline for our experiments. 
We chose this baseline as a representative of the traditional entity normalization methods that provides competitive results based on an ensemble of such models.  

\noindent {\bf BioBERT.} 
To compare our method with a representative of text embedding approaches, we used the embedding generated by the recently released BioBERT~\citep{lee2019biobert} (Bidirectional Encoder Representations from Transformers for Biomedical Text Mining) model which extends the BERT~\citep{devlin2018bert} approach. BioBERT is a domain specific language representation model pre-trained on large-scale biomedical corpora that can effectively capture knowledge from a large amount of biomedical texts with minimal task-specific architecture modifications. BioBERT outperforms traditional models in biomedical named entity recognition, biomedical relation extraction, and biomedical question answering.
We used the BioBERT framework with pre-trained weights released by the original authors of the paper, in a similar process to our approach; we first embed all the entity names of the reference set and then find the closest embedding to the query name in that embedding space. 

\noindent {\bf DeepMatcher.}
\citet{mudgal2018deep} recently studied the application of deep learning architectures on entity matching in a general setting where the task is matching tuples (potentially having multiple fields) in different tables. DeepMatcher outperforms traditional entity matching frameworks in textual and noisy settings.
We use DeepMatcher as a representative baseline for deep learning methods specific to entity normalization.

We used the implementation published by the authors to perform our experiments. 
We used DeepMatcher with tuples containing only one field; the entity mention. We train DeepMatcher with the same initial pairs we use to train our model, and follow a common-word-based blocking technique recommended in their implementation to pre-process our data.
DeepMatcher does not perform hard negative mining during its training, and the blocking is performed prior to the matching process in contrast to our framework. 

\subsection{Results}
Table~\ref{tab:results} shows the comparative results of our method (i.e., \NSR) with other methods. 
We submit every query name in the BioCreative datasets to all systems, and retrieve the top \textit{k} most probable IDs from each of them. 
We then find out if the correct ID (provided in the BioCreative dataset as labels) is present in the top~\textit{k} retrieved results (i.e., \textit{Hits@k}) for several values of \textit{k}. 
Our method outperforms the baselines in almost all settings. 
Chemical names are generally harder to normalize due to more sensitivity to parenthesis, commas, and dashes, but our method produces significantly better results.

\begin{table}[tbp]
\centering
\caption{Hits@k on BioCreative train dataset (BioC1) and test dataset (BioC2) datasets mapped to Uniprot and ChEBI reference sets.}\label{tab:results}
    \begin{tabularx}{\hsize}{@{}l l l X X X X@{}}
    \toprule
    \textbf{Reference($\mathcal{R}$)} & \textbf{Dataset} & \textbf{Method} & \textbf{H@1} & \textbf{H@3} & \textbf{H@5} & \textbf{H@10} \\
    \midrule

    \multirow{8}{*}{\textbf{UniProt}}
    &\multirow{4}{*}{BioC1} 
    & DeepMatcher & 0.697 & 0.728 & 0.739 & 0.744\\
    & & BioBERT & 0.729 & 0.761 & 0.779 & 0.808\\
    & & USC--ISI & 0.814 & 0.864 & 0.875 & 0.885\\
    & &\textbf{\NSR} & \textbf{0.833} & \textbf{0.869} & \textbf{0.886} & \textbf{0.894}\\
    \cline{3-7}
    &\multirow{4}{*}{BioC2} 
    & DeepMatcher & 0.767 & 0.792 & 0.803 & 0.814\\
    & & BioBERT & 0.801 & 0.827 & 0.827 & 0.840\\
    & & USC--ISI & 0.841 & 0.888 & 0.904 & 0.919 \\
    & &\textbf{\NSR} & \textbf{0.861} & 0.888 & 0.904 & \textbf{0.930}\\
    \midrule
    \multirow{8}{*}{\textbf{ChEBI}}
    &\multirow{4}{*}{BioC1} 
    & DeepMatcher & 0.288 & 0.363 & 0.397 & 0.419\\
    & & BioBERT & 0.360 & 0.473 & 0.499 & 0.524\\
    & & USC--ISI & 0.418 & 0.451 & 0.460 & 0.468\\
    & &\textbf{\NSR} & \textbf{0.505} & \textbf{0.537} & \textbf{0.554} & \textbf{0.574}\\
    \cline{3-7}
    &\multirow{4}{*}{BioC2} 
    & DeepMatcher & 0.373 & 0.463 & 0.491 & 0.517\\
    & & BioBERT & 0.422 & 0.558 & 0.577 & 0.596\\
    & & USC--ISI & 0.444 & 0.472 & 0.480 & 0.491 \\    
    & &\textbf{\NSR} & \textbf{0.578} & \textbf{0.608} & \textbf{0.624} & \textbf{0.641}\\
    \bottomrule
    \end{tabularx}%
\end{table}

\begin{figure*}[t]
    \centering
    \subfloat[S6K]{{\includegraphics[width=0.25\textwidth,trim={1.2cm 1.2cm 1.2cm 1.2cm},clip]{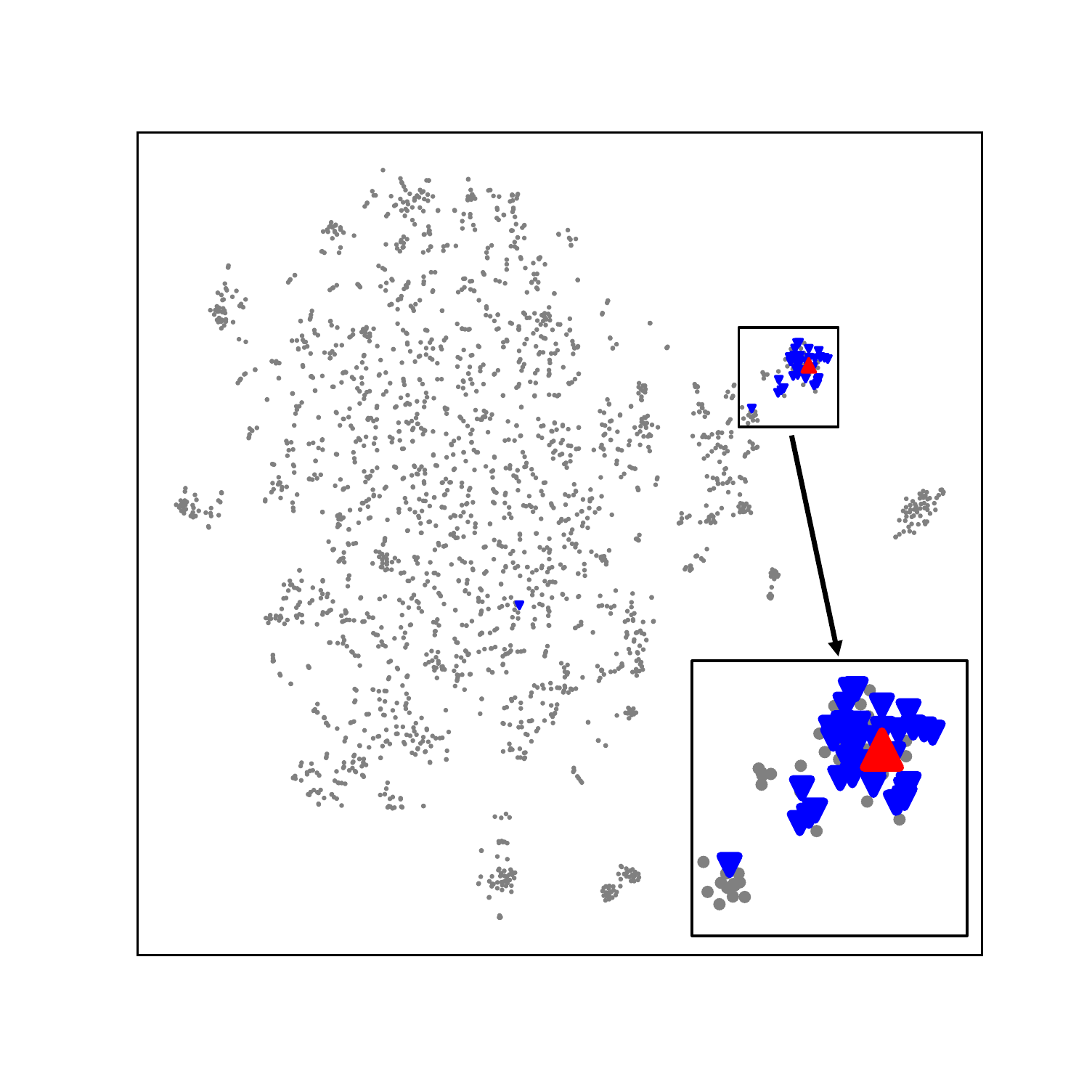}}}%
    \subfloat[PLC$\gamma$2]{{\includegraphics[width=0.25\textwidth,trim={1.2cm 1.2cm 1.2cm 1.2cm},clip]{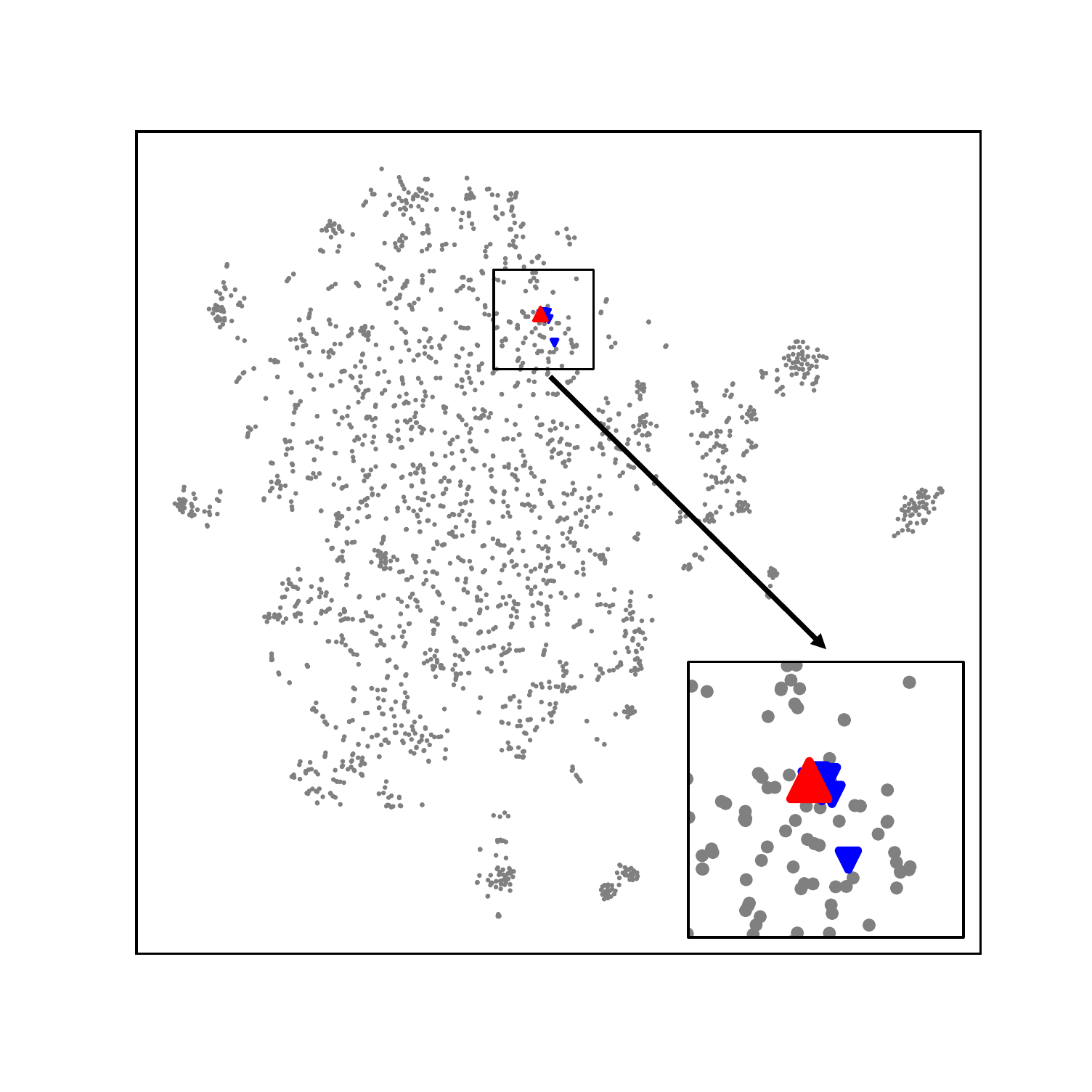}}}%
    \subfloat[IKK$\epsilon$]{{\includegraphics[width=0.25\textwidth,trim={1.2cm 1.2cm 1.2cm 1.2cm},clip]{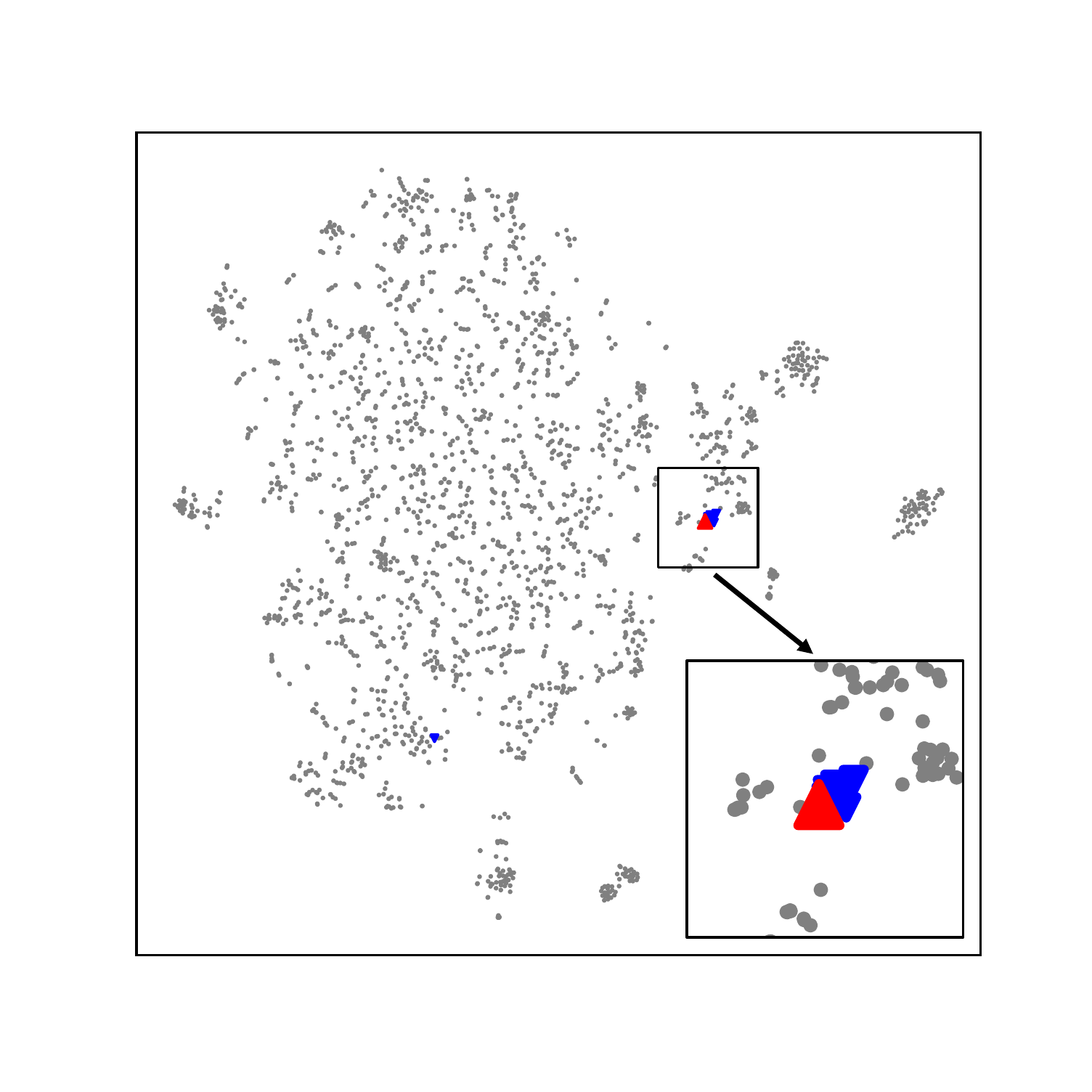}}}%
    \subfloat[H3]{{\includegraphics[width=0.25\textwidth,trim={1.2cm 1.2cm 1.2cm 1.2cm},clip]{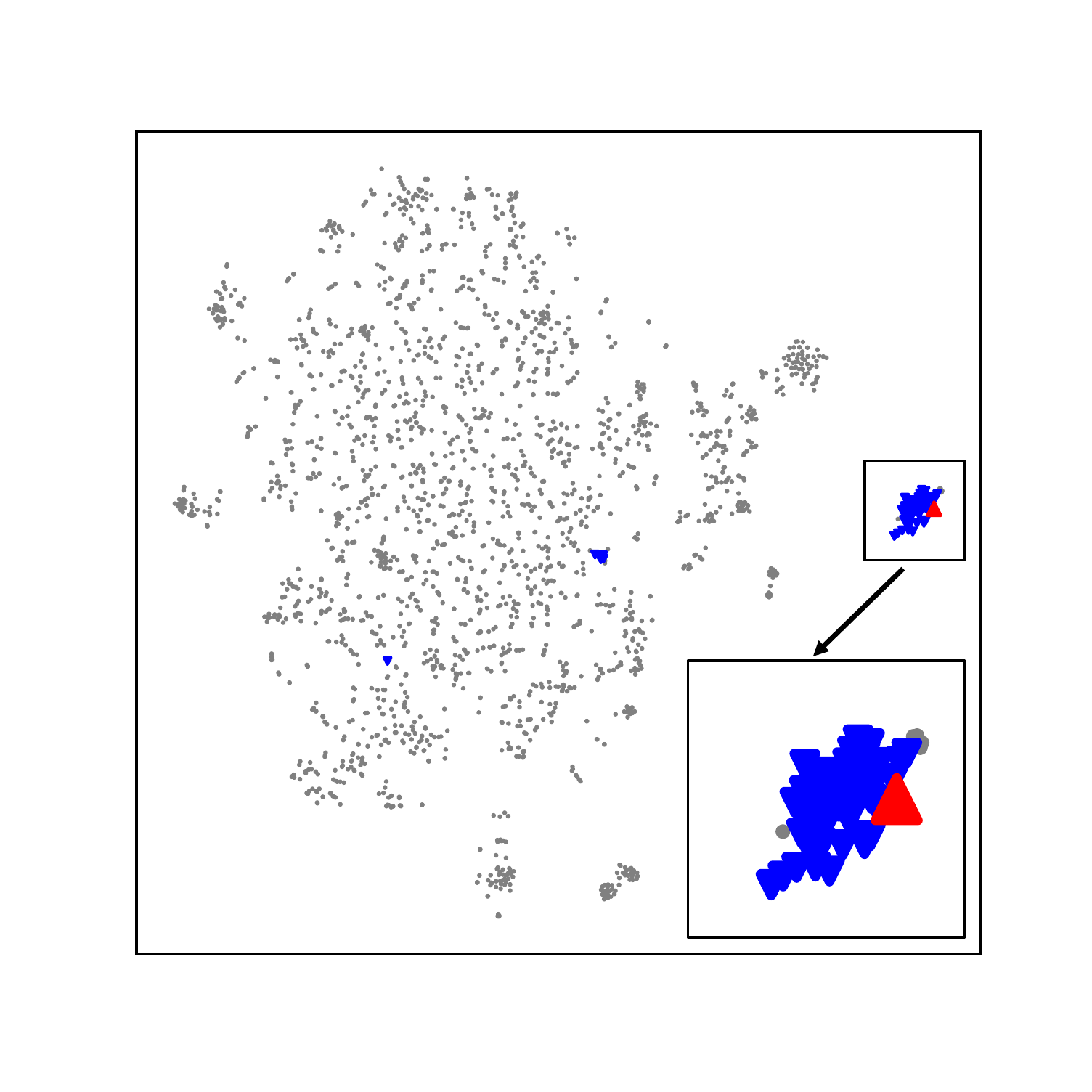}}}%
        
    \caption{tSNE representation of the example UniPort query entities shown in Table~\ref{tab:uniprot_queries}. Queries are red triangle and correct responses are blue. A sample of a thousand names from the reference set is shown with light grey dots to represent the embedding space. The bottom right insets show a zoomed version of the correct names clustered around the query name. (Best viewed in color) }%
    \label{fig:uniprot_tsne}%
\end{figure*}

Furthermore, Table~\ref{tab:uniprot_queries} and the corresponding Figure~\ref{fig:uniprot_tsne} show example protein name queries mapped to the UniProt reference set and the retrieved canonical names.
Note that \textit{none of the query names exist in the UniProt} reference set in the form provided as the query. 
Table~\ref{tab:uniprot_queries} shows not only the syntactic variations being captured by our method in the Top 10 responses, but the semantically equivalent names are included as well. These responses can have a significantly large string distance with the query name. 
e.g., 
(\textit{S6K}$\longrightarrow$\textit{52 kDa ribosomal protein S6 kinase}), (\textit{PLC$\gamma$2}$\longrightarrow$\textit{Phospholipase C-gamma-2}), (\textit{IKK$\epsilon$}$\longrightarrow$\textit{I-kappa-B kinase epsilon}), and (\textit{H3}$\longrightarrow$\textit{Histone H3/a}). 

\begin{table*}[t]
\centering
\caption{UniProt sample queries and top-10 responses. The correct entities are indicated with a bold font and an asterisk. None of the queries have an exact string match in UniProt, and the lists include syntactically far correct responses.}\label{tab:uniprot_queries}
    \begin{tabularx}{\hsize}{@{}X X X X@{}}
    \toprule
    \textbf{S6K} & \textbf{PLC$\gamma$2} & \textbf{IKK$\epsilon$} & \textbf{H3}\\
    \midrule
    \pbox[t]{0.25\textwidth}{
    \scriptsize - \textbf{p70-S6K 1}*\\
    \scriptsize - p90-RSK 6\\
    \scriptsize - \textbf{S6K1}*\\
    \scriptsize - \textbf{p70 S6KA}*\\
    \scriptsize - S6K-beta\\
    \scriptsize - p70 S6KB\\
    \scriptsize - 90 kDa ribosomal\\ $~~$ protein S6 kinase 6\\
    \scriptsize - 90 kDa ribosomal\\ $~~$ protein S6 kinase 5\\
    \scriptsize - \textbf{52 kDa ribosomal}\\ $~~$ \textbf{protein S6 kinase}*\\
    \scriptsize - RPS6KA6\\
    ~}
    &
    \pbox[t]{0.25\textwidth}{
    \scriptsize - \textbf{PLC-gamma-2}*\\
    \scriptsize - PLC-gamma-1\\
    \scriptsize - \textbf{PLCG2}*\\
    \scriptsize - \textbf{Phospholipase}\\ $~~$ \textbf{C-gamma-2}*\\
    \scriptsize - Phospholipase \\ $~~$ C-gamma-1\\
    \scriptsize - PLC\\
    \scriptsize - PLCG1\\
    \scriptsize - \textbf{Phosphoinositide} \\ $~~$ \textbf{phospholipase} \\ $~~$ \textbf{C-gamma-2}*\\
    \scriptsize - \textbf{PLC-IV}*\\
    \scriptsize - PLCB\\
    ~}
    & 
    \pbox[t]{0.25\textwidth}{
    \scriptsize - \textbf{IKK-epsilon}*\\
    \scriptsize - \textbf{IKKE}*\\
    \scriptsize - \textbf{I-kappa-B kinase}\\ $~~$ \textbf{epsilon}*\\
    \scriptsize - \textbf{IkBKE}*\\
    \scriptsize - \textbf{IKBKE}*\\
    \scriptsize - IKBE\\
    \scriptsize - IK1\\
    \scriptsize - IK1\\
    \scriptsize - IKKG\\
    \scriptsize - INKA1\\
    ~}
    &
    \pbox[t]{0.25\textwidth}{
    \scriptsize - \textbf{Histone H3/a}*\\
    \scriptsize - \textbf{Histone H3/o}*\\
    \scriptsize - \textbf{Histone H3/m}*\\
    \scriptsize - \textbf{Histone H3/b}*\\
    \scriptsize - \textbf{Histone H3/f}*\\
    \scriptsize - \textbf{HIST1H3C}*\\
    \scriptsize - \textbf{Histone H3/k}*\\
    \scriptsize - \textbf{Histone H3/i}*\\
    \scriptsize - \textbf{HIST1H3G}*\\
    \scriptsize - \textbf{Histone H3/d}*\\
    ~}
    \\
    \bottomrule
    \end{tabularx}%
\end{table*}

Figure~\ref{fig:uniprot_tsne} sheds more light to the embedding space and highlights the same four query names and the names corresponding to the correct entities in the UniProt reference set. As shown in this figure most of the correct responses (in blue) are clustered around the query name (in red).

The retrieval time of the baseline methods are in the order of a few minutes. \NSR\ relies on the approximate nearest neighbors architecture and provides highly competitive retrieval performance in the order of seconds. 
The study reported on \citep{bernhardssonannoy} for approximate nearest neighbors architectures applies to our method as well. 
\section{Discussion}
In this paper, we proposed a general deep neural network based framework for entity normalization.
We showed how to encode semantic information hidden in a reference set, and how to incorporate potential syntactic variations in the numeric embedding space via training-pair generation.
In this process we showed how contrastive loss can be used with non-binary labels to capture uncertainty. 
We further introduced a dynamic hard negative sampling method to refine the embeddings. 
Finally, by transforming the traditional task of entity normalization to a standard k-nearest neighbors problem in a numerical space, we showed how to employ a scalable representation for fast retrievals that is applicable in real-world scenarios without the need of traditional entity blocking methods. 
By eliminating the need for blocking as a pre-processing step, we can consider matches that are syntactically different but semantically relevant, which is not easily achievable via traditional entity normalization methods. 

In our preliminary analysis, we experimented with different selection methods in the k-nearest neighbors retrieval process such as a \textit{top-k majority vote} schema, but did not find them significantly effective in our setting.
We also experimented with different soft labeling methods to dynamically re-rank the results such as \textit{soft re-labeling the k-nearest neighbors}, but did not see much improvements to the overall performance.
While currently highly effective, our method could benefit from improving some of its components in future research. 
We are also considering combining our approach with other embedding and collective reasoning methods to gain further potential performance improvements. 

\subsection*{Acknowledgments}
This work was supported in part by DARPA Big Mechanism program under contract number W911NF-14-1-0364.

\bibliographystyle{unsrtnat}
\bibliography{references}

\begin{thebibliography}{38}
\providecommand{\natexlab}[1]{#1}
\providecommand{\url}[1]{\texttt{#1}}
\expandafter\ifx\csname urlstyle\endcsname\relax
  \providecommand{\doi}[1]{doi: #1}\else
  \providecommand{\doi}{doi: \begingroup \urlstyle{rm}\Url}\fi

\bibitem[Yadav and Bethard(2018)]{yadav2018survey}
Vikas Yadav and Steven Bethard.
\newblock A survey on recent advances in named entity recognition from deep
  learning models.
\newblock In \emph{Proceedings of the 27th International Conference on
  Computational Linguistics}, 2018.

\bibitem[Mathew et~al.(2019)Mathew, Fakhraei, and Ambite]{mathew2019biomedical}
Joel Mathew, Shobeir Fakhraei, and Jos{\'e}~Luis Ambite.
\newblock Biomedical named entity recognition via reference-set augmented
  bootstrapping.
\newblock \emph{ICML Workshop on Computational Biology}, 2019.

\bibitem[Papadakis et~al.(2016)Papadakis, Svirsky, Gal, and
  Palpanas]{papadakis2016comparative}
George Papadakis, Jonathan Svirsky, Avigdor Gal, and Themis Palpanas.
\newblock Comparative analysis of approximate blocking techniques for entity
  resolution.
\newblock \emph{Proceedings of the VLDB Endowment}, 2016.

\bibitem[Christen(2012{\natexlab{a}})]{christen2012data}
Peter Christen.
\newblock \emph{Data matching: concepts and techniques for record linkage,
  entity resolution, and duplicate detection}.
\newblock Springer Science \& Business Media, 2012{\natexlab{a}}.

\bibitem[Koudas et~al.(2006)Koudas, Sarawagi, and Srivastava]{koudas2006record}
Nick Koudas, Sunita Sarawagi, and Divesh Srivastava.
\newblock Record linkage: similarity measures and algorithms.
\newblock In \emph{Proceedings of the 2006 ACM SIGMOD international conference
  on Management of data}, 2006.

\bibitem[Elmagarmid et~al.(2007)Elmagarmid, Ipeirotis, and
  Verykios]{elmagarmid2007duplicate}
Ahmed~K Elmagarmid, Panagiotis~G Ipeirotis, and Vassilios~S Verykios.
\newblock Duplicate record detection: A survey.
\newblock \emph{IEEE TKDE}, 2007.

\bibitem[Getoor and Machanavajjhala(2012)]{getoor2012entity}
Lise Getoor and Ashwin Machanavajjhala.
\newblock Entity resolution: theory, practice \& open challenges.
\newblock \emph{Proceedings of the VLDB Endowment}, 2012.

\bibitem[Leaman and Lu(2016)]{leaman2016taggerone}
Robert Leaman and Zhiyong Lu.
\newblock {TaggerOne}: joint named entity recognition and normalization with
  semi-markov models.
\newblock \emph{Bioinformatics}, 2016.

\bibitem[Cheatham and Hitzler(2013)]{cheatham2013string}
Michelle Cheatham and Pascal Hitzler.
\newblock String similarity metrics for ontology alignment.
\newblock In \emph{International Semantic Web Conference}, 2013.

\bibitem[Mikolov et~al.(2013)Mikolov, Sutskever, Chen, Corrado, and
  Dean]{mikolov2013distributed}
Tomas Mikolov, Ilya Sutskever, Kai Chen, Greg~S Corrado, and Jeff Dean.
\newblock Distributed representations of words and phrases and their
  compositionality.
\newblock In \emph{Advances in neural information processing systems}, 2013.

\bibitem[Pennington et~al.(2014)Pennington, Socher, and
  Manning]{pennington2014glove}
Jeffrey Pennington, Richard Socher, and Christopher Manning.
\newblock Glove: Global vectors for word representation.
\newblock In \emph{Proceedings of the 2014 conference on empirical methods in
  natural language processing (EMNLP)}, 2014.

\bibitem[Peters et~al.(2018)Peters, Neumann, Iyyer, Gardner, Clark, Lee, and
  Zettlemoyer]{Peters:2018}
Matthew~E. Peters, Mark Neumann, Mohit Iyyer, Matt Gardner, Christopher Clark,
  Kenton Lee, and Luke Zettlemoyer.
\newblock Deep contextualized word representations.
\newblock In \emph{Proc. of NAACL}, 2018.

\bibitem[Devlin et~al.(2018)Devlin, Chang, Lee, and Toutanova]{devlin2018bert}
Jacob Devlin, Ming-Wei Chang, Kenton Lee, and Kristina Toutanova.
\newblock Bert: Pre-training of deep bidirectional transformers for language
  understanding.
\newblock \emph{arXiv preprint arXiv:1810.04805}, 2018.

\bibitem[Neculoiu et~al.(2016)Neculoiu, Versteegh, and
  Rotaru]{neculoiu2016learning}
Paul Neculoiu, Maarten Versteegh, and Mihai Rotaru.
\newblock Learning text similarity with siamese recurrent networks.
\newblock In \emph{Proceedings The 1st Workshop on Representation Learning for
  NLP}, 2016.

\bibitem[Taigman et~al.(2014)Taigman, Yang, Ranzato, and
  Wolf]{taigman2014deepface}
Yaniv Taigman, Ming Yang, Marc'Aurelio Ranzato, and Lior Wolf.
\newblock Deepface: Closing the gap to human-level performance in face
  verification.
\newblock In \emph{Proceedings of the IEEE conference on computer vision and
  pattern recognition}, 2014.

\bibitem[Hadsell et~al.(2006)Hadsell, Chopra, and
  LeCun]{hadsell2006dimensionality}
R~Hadsell, S~Chopra, and Y~LeCun.
\newblock Dimensionality reduction by learning an invariant mapping.
\newblock In \emph{Computer Vision and Pattern Recognition, 2006 IEEE Computer
  Society Conference on}, 2006.

\bibitem[Joty and Tang(2018)]{joty2018distributed}
Muhammad Ebraheem Saravanan Thirumuruganathan~Shafiq Joty and Mourad
  Ouzzani~Nan Tang.
\newblock Distributed representations of tuples for entity resolution.
\newblock \emph{Proceedings of the VLDB Endowment}, 2018.

\bibitem[Mudgal et~al.(2018)Mudgal, Li, Rekatsinas, Doan, Park, Krishnan, Deep,
  Arcaute, and Raghavendra]{mudgal2018deep}
Sidharth Mudgal, Han Li, Theodoros Rekatsinas, AnHai Doan, Youngchoon Park,
  Ganesh Krishnan, Rohit Deep, Esteban Arcaute, and Vijay Raghavendra.
\newblock Deep learning for entity matching: A design space exploration.
\newblock In \emph{Proceedings of the 2018 International Conference on
  Management of Data}, 2018.

\bibitem[Michelson and Knoblock(2006)]{michelson2006learning}
Matthew Michelson and Craig~A Knoblock.
\newblock Learning blocking schemes for record linkage.
\newblock In \emph{AAAI}, 2006.

\bibitem[Christen(2012{\natexlab{b}})]{christen2012survey}
Peter Christen.
\newblock A survey of indexing techniques for scalable record linkage and
  deduplication.
\newblock \emph{IEEE TKDE}, 2012{\natexlab{b}}.

\bibitem[Kang et~al.(2012)Kang, Singh, Afzal, van Mulligen, and
  Kors]{kang2012using}
Ning Kang, Bharat Singh, Zubair Afzal, Erik~M van Mulligen, and Jan~A Kors.
\newblock Using rule-based natural language processing to improve disease
  normalization in biomedical text.
\newblock \emph{JAMIA}, 2012.

\bibitem[Leaman et~al.(2013)Leaman, Islamaj~Do{\u{g}}an, and
  Lu]{leaman2013dnorm}
Robert Leaman, Rezarta Islamaj~Do{\u{g}}an, and Zhiyong Lu.
\newblock Dnorm: disease name normalization with pairwise learning to rank.
\newblock \emph{Bioinformatics}, 2013.

\bibitem[Cohen et~al.(2003)Cohen, Ravikumar, and Fienberg]{cohen2003comparison}
William Cohen, Pradeep Ravikumar, and Stephen Fienberg.
\newblock A comparison of string metrics for matching names and records.
\newblock In \emph{KDD workshop on data cleaning and object consolidation},
  2003.

\bibitem[Chen et~al.(2018)Chen, Perozzi, Hu, and Skiena]{chen2018harp}
Haochen Chen, Bryan Perozzi, Yifan Hu, and Steven Skiena.
\newblock {HARP}: Hierarchical representation learning for networks.
\newblock 2018.

\bibitem[Kotnis and Nastase(2017)]{kotnis2017analysis}
Bhushan Kotnis and Vivi Nastase.
\newblock Analysis of the impact of negative sampling on link prediction in
  knowledge graphs.
\newblock In \emph{WSDM 1st Workshop on Knowledge Base Construction, Reasoning
  and Mining (KBCOM)}, 2017.

\bibitem[Shrivastava et~al.(2016)Shrivastava, Gupta, and
  Girshick]{shrivastava2016training}
Abhinav Shrivastava, Abhinav Gupta, and Ross Girshick.
\newblock Training region-based object detectors with online hard example
  mining.
\newblock In \emph{Proceedings of the IEEE Conference on Computer Vision and
  Pattern Recognition}, 2016.

\bibitem[Ponomarenko et~al.(2014)Ponomarenko, Avrelin, Naidan, and
  Boytsov]{ponomarenko2014comparative}
Alexander Ponomarenko, Nikita Avrelin, Bilegsaikhan Naidan, and Leonid Boytsov.
\newblock Comparative analysis of data structures for approximate nearest
  neighbor search.
\newblock \emph{Data Analytics}, 2014.

\bibitem[Rastegari et~al.(2013)Rastegari, Choi, Fakhraei, Hal, and
  Davis]{rastegari2013predictable}
Mohammad Rastegari, Jonghyun Choi, Shobeir Fakhraei, Daume Hal, and Larry
  Davis.
\newblock Predictable dual-view hashing.
\newblock In \emph{Int. Conference on Machine Learning (ICML)}, 2013.

\bibitem[ber(2019)]{bernhardssonannoy}
Annoy (approximate nearest neighbors oh yeah).
\newblock \url{https://github.com/spotify/annoy}, 2019.

\bibitem[Bachrach et~al.(2014)Bachrach, Finkelstein, Gilad-Bachrach, Katzir,
  Koenigstein, Nice, and Paquet]{bachrach2014speeding}
Yoram Bachrach, Yehuda Finkelstein, Ran Gilad-Bachrach, Liran Katzir, Noam
  Koenigstein, Nir Nice, and Ulrich Paquet.
\newblock Speeding up the xbox recommender system using a euclidean
  transformation for inner-product spaces.
\newblock In \emph{Proceedings of the 8th ACM Conference on Recommender
  systems}, 2014.

\bibitem[Naidan and Boytsov(2015)]{naidan2015non}
Bilegsaikhan Naidan and Leonid Boytsov.
\newblock Non-metric space library manual.
\newblock \emph{arXiv preprint arXiv:1508.05470}, 2015.

\bibitem[Arighi et~al.(2017)Arighi, Hirschman, Lemberger, Bayer, Liechti,
  Comeau, and Wu]{arighi2017bio}
Cecilia Arighi, Lynette Hirschman, Thomas Lemberger, Samuel Bayer, Robin
  Liechti, Donald Comeau, and Cathy Wu.
\newblock Bio-id track overview.
\newblock In \emph{Proceedings of the BioCreative VI Workshop}, 2017.

\bibitem[Apweiler et~al.(2004)Apweiler, Bairoch, Wu, Barker, Boeckmann, Ferro,
  Gasteiger, Huang, Lopez, Magrane, et~al.]{apweiler2004uniprot}
Rolf Apweiler, Amos Bairoch, Cathy~H Wu, Winona~C Barker, Brigitte Boeckmann,
  Serenella Ferro, Elisabeth Gasteiger, Hongzhan Huang, Rodrigo Lopez, Michele
  Magrane, et~al.
\newblock Uniprot: the universal protein knowledgebase.
\newblock \emph{Nucleic acids research}, 2004.

\bibitem[Hastings et~al.(2015)Hastings, Owen, Dekker, Ennis, Kale,
  Muthukrishnan, Turner, Swainston, Mendes, and Steinbeck]{hastings2015chebi}
Janna Hastings, Gareth Owen, Adriano Dekker, Marcus Ennis, Namrata Kale,
  Venkatesh Muthukrishnan, Steve Turner, Neil Swainston, Pedro Mendes, and
  Christoph Steinbeck.
\newblock Chebi in 2016: Improved services and an expanding collection of
  metabolites.
\newblock \emph{Nucleic acids research}, 2015.

\bibitem[amb(2018)]{ambite2017}
{University of Southern California - Information Science Institute Entity
  Grounding System}.
\newblock \url{http://dna.isi.edu:7100/}, 2018.

\bibitem[Jurczyk et~al.(2008)Jurczyk, Lu, Xiong, Cragan, and
  Correa]{jurczyk2008fril}
Pawel Jurczyk, James~J Lu, Li~Xiong, Janet~D Cragan, and Adolfo Correa.
\newblock Fril: a tool for comparative record linkage.
\newblock In \emph{American Medical Informatics Association (AMIA) annual
  symposium proceedings}, 2008.

\bibitem[Bia{\l}ecki et~al.(2012)Bia{\l}ecki, Muir, and
  Ingersoll]{bialecki2012apache}
Andrzej Bia{\l}ecki, Robert Muir, and Grant Ingersoll.
\newblock Apache lucene 4.
\newblock In \emph{SIGIR 2012 workshop on open source information retrieval},
  2012.

\bibitem[Lee et~al.(2019)Lee, Yoon, Kim, Kim, Kim, So, and
  Kang]{lee2019biobert}
Jinhyuk Lee, Wonjin Yoon, Sungdong Kim, Donghyeon Kim, Sunkyu Kim, Chan~Ho So,
  and Jaewoo Kang.
\newblock {BioBERT}: pre-trained biomedical language representation model for
  biomedical text mining.
\newblock \emph{arXiv preprint arXiv:1901.08746}, 2019.

\end{thebibliography}

\end{document}